\documentclass[superscriptaddress,pra, amsmath,amssymb,preprint,showkeys,showpacs]{revtex4-1}

\usepackage{graphicx}
\usepackage{graphics}
\usepackage{amssymb}
\usepackage{amsmath,bm}
\usepackage{float}
\usepackage[usenames,dvipsnames]{color}
\usepackage[export]{adjustbox}
\usepackage{hyperref}
\hypersetup{
    colorlinks=true,
    linkcolor=blue,
    citecolor=blue,
    filecolor=cyan,      
    urlcolor=cyan,
}
\begin{document}
\title{Spin-optomechanical coupling between light and a nanofiber torsional mode}


\author{Eliot F. Fenton}

\affiliation{Joint Quantum Institute and Department of Physics, University of Maryland, College Park, MD 20742, USA.}

\author{Adnan Khan}
\affiliation{Joint Quantum Institute and Department of Physics, University of Maryland, College Park, MD 20742, USA.}

\author{Pablo Solano}
\affiliation{Joint Quantum Institute and Department of Physics, University of Maryland, College Park, MD 20742, USA.}

\author{Luis A. Orozco}
\affiliation{Joint Quantum Institute and Department of Physics,
University of Maryland, College Park, MD 20742, USA.}

\author{Fredrik K. Fatemi}
 \email{Corresponding author email:  fredrik.k.fatemi.civ@mail.mil}
\affiliation{Army Research Laboratory, Adelphi, MD 20783, USA}

%
%




\date{\today}

\begin{abstract}
Light that carries linear or angular momentum can interact with a mechanical object giving rise to optomechanical effects. In particular, a photon transfers its intrinsic angular momentum to an object when the object either absorbs the photon or changes the photon polarization, as in an action/reaction force pair.
Here, we present the implementation of light-induced selective resonant driving of the torsional mechanical modes of a single-mode tapered optical nanofiber. The nanofiber torsional mode spectrum is characterized by polarimetry, showing narrow natural resonances (Q$\approx$ 2,000). By sending amplitude modulated light through the nanofiber, we resonantly drive individual torsional modes as a function of the light polarization. By varying the input polarization to the fiber, we find the largest amplification of a mechanical oscillation ($>$35 dB) is observed when driving the system with light containing longitudinal spin on the nanofiber waist. These results present optical nanofibers as a platform suitable for quantum spin-optomechanics experiments.
\end{abstract}


\maketitle


The small amount of angular momentum carried by light usually produces an imperceptible torque on mechanical objects, making it challenging to observe. The pioneering experiment of Beth in 1936 \cite{Berth1936} coupled the intrinsic angular momentum of a circularly polarized beam of light to a quartz plate and measured the rotation of the plate through a torsional pendulum.  Since then, spin angular momentum has been transferred in a number of systems, including  birefringent \cite{Friese, Arias} and absorptive~\cite{He1995,Friese1996} microparticles.  Optomechanical torque can be amplified by bringing the mechanical system to the nanoscale and driving it at its natural resonant frequency ~\cite{He2016}.  In that work, employing a short (~10-um-long) rectangular nanobeam, the different propagation constants of the $TE$ and $TM$ modes yield a strong geometric birefringence through which spin angular momentum can be transferred.

We use optical nanofibers (ONF) \cite{Solano2017d} as a platform for spin-optomechanics for the work in this letter.  These nanofibers have cylindrical symmetry and hence very low intrinsic birefringence, such that the transfer of spin angular momentum relies on the absorption, albeit low, of the waveguide. The torsional mechanics of ONFs have been studied in Ref. \cite{Wuttke2013} through direct mechanical excitation, exhibiting intrinsically narrow torsional modes, but excitation through optical means has not been explored to the best of our knowledge.  However, the strong light confinement afforded by ONFs make them good candidates for studying optical-to-mechanical coupling of angular momentum.  The stress produced by the torsion on the nanofiber affects its index of refraction creating a birefringent medium. This torque-induced birefringence can be directly measured through polarimetry of a guided field, giving information of the torsional dynamics of the system.

We selectively excite and amplify a torsional mode of the nanofiber by more than 35 dB by sending amplitude modulated (AM) light. Excitation is maximized for circularly-polarized light modulated at the mechanical resonant frequency. We show control of the amplitude and phase of the driven system. The mechanical excitation and optical readout are both performed through an ultrahigh transmission ONF \cite{Hoffman2014a}. Our results present ONFs as a platform suitable for spin-optomechanical experiments. Their compatibility with cryogenic systems to reduce phonon occupation, and large coupling with the surrounding media through the evanescent field, might play an important role in spin-optomechanical applications reaching single photon sensitivity.

\section{System}

\begin{figure}[h]
\centering
\includegraphics[width=0.7\linewidth]{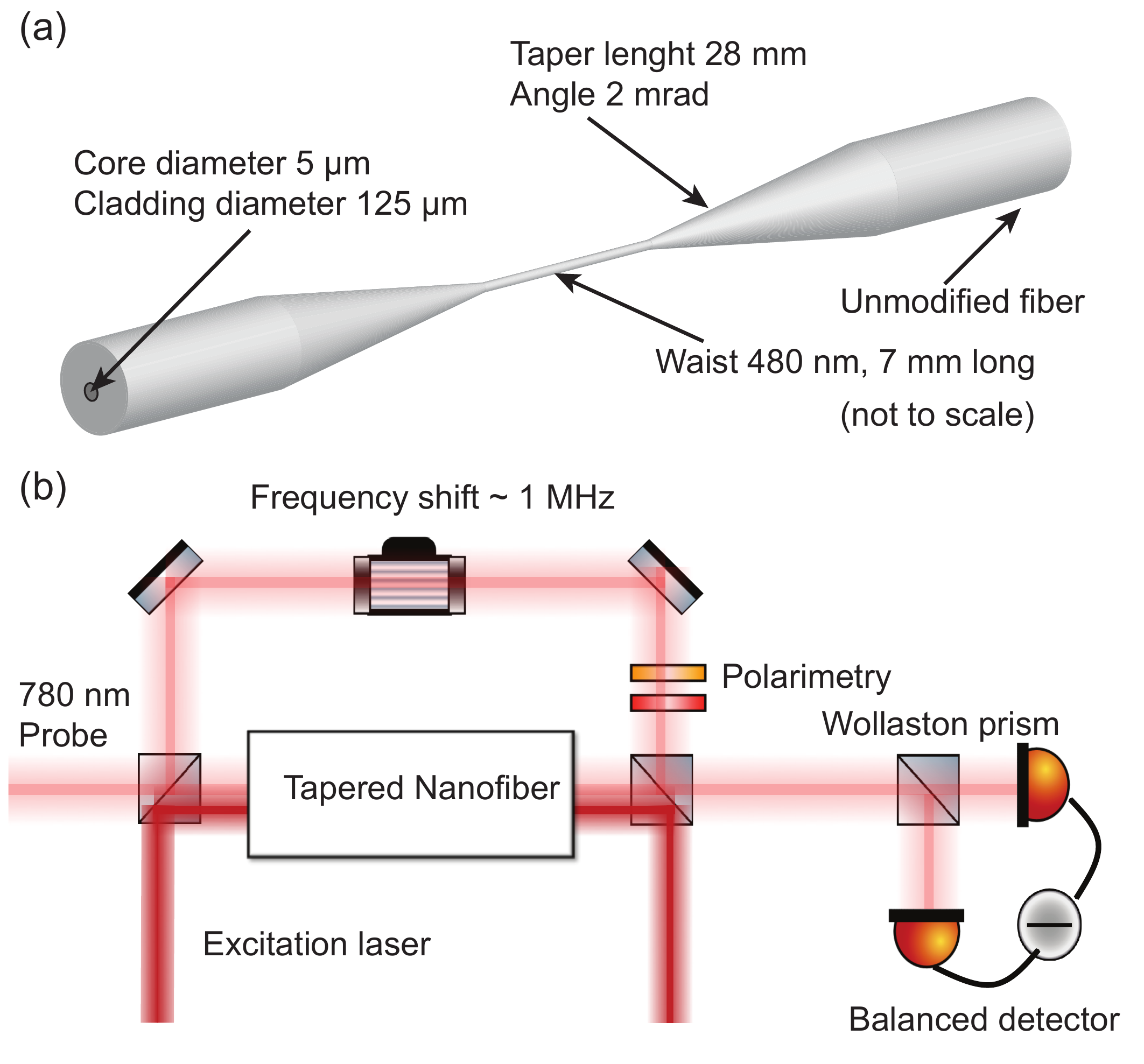}
\caption{System with fiber and detection. (a) Tapered optical nanofiber showing the five sections with the sizes of one the actual fiber used for the measurements (ONF 1). (b) Heterodyne measurement scheme. AOMs shift the frequency of the 780 nm probe beam that does not pass through the nanofiber. The dark red beam represents either the driving or heating 1064 nm beam that can be turned on or off as needed.}
\label{fig:ExpSetup}
\end{figure}

Figure \ref{fig:ExpSetup} shows the nanofiber and the schematic of the experimental apparatus. The ONF, with its five sections as marked in Fig. \ref{fig:ExpSetup}(a), is produced via the flame brushing technique~\cite{Hoffman2014a,Birks1992}. A hydrogen-oxygen flame acts as a local heat source to soften a single-mode fiber, the ends of which are pulled with computer-controlled linear motors. This method reliably produces fibers of subwavelength diameters with transmission of the fundamental mode above 99\% and as high as 99.95\%, allowing them to sustain powers of hundreds of milliwatts in high vacuum~\cite{Hoffman2014a}. At these dimensions, the ONF waist is dominated by the cladding and can be considered as a simple dielectric of index $n_\mathrm{ONF}$ = $n_\mathrm{clad} \approx 1.45$, surrounded by $n$ = 1.0, as the original fiber core becomes negligible. 

This paper presents results using two different tapered nanofibers. The first nanofiber (ONF 1) is  from a single mode optical fiber (Fibercore SM800) and tapered down to $480\pm 40$ nm diameter with a 1\% uniformity over a waist length of 7 mm, and a transmission greater than 95\%. The uncertainties are estimated based on our fiber-pulling reproducibility. Each taper section connecting the standard fiber to the input and output side of ONF 1 is about 28 mm long. The tapers which have first a linear and then an exponential section joining the waist, connect the standard fiber on the input and output side to the ONF. The taper angle of $\Omega=2$ mrad ensures adiabatic propagation of the fundamental $HE_{11}$ mode ~\cite{Ravets2013a}. The second fiber (ONF 2) is  from a few-mode optical fiber (Fibercore SM1500) and it has a 10-mm-long, $780\pm 60$-nm-diameter waist, with taper angle of $\Omega=1$ mrad. For 780-nm light, this fiber supports the $LP_{01}$ and $LP_{11}$ families of modes, both in the unmodified section and in the waist. 

The ONFs are glued (EPO-TEK OG116-31) to a titanium U-shaped fixture for stability, and mounted in ultra-high vacuum (UHV). In particular, ONF 1 has been under UHV (better than 10$^{-9}$ Torr) for more than three years, and has been used for numerous experiments with Rb atoms in its vicinity \cite{Grover2015,Solano2017a,Solano2017b,Solano2017c}. 
 
Mechanical motion in the ONF causes strain-induced birefringence that rotates the polarization of light propagating through it~\cite{Wuttke2014}. To detect torsional modes, we monitored these oscillating strain-induced polarization rotations of a probe beam that passed through the ONF. Because the frequencies of the modes shifted with high optical power in the ONF, due to heating, accurate measurement of the unperturbed torsional mode frequencies required a low-power ($<$1 $\mu$W) probe beam. 

We use a heterodyne detection scheme to study ONF 1 (Fig. \ref{fig:ExpSetup})(b). We split a 780 nm wavelength laser beam (Toptica TA pro) into a probe beam and local oscillator (LO) at a 50/50 beam splitter. We shifted the frequency of the probe beam by +199 MHz using a double-pass acousto-optic modulator (AOM) at 99.5 MHz and then by -200 MHz using a single-pass AOM for a total frequency shift of 1 MHz down relative to the LO. We further decreased the probe beam power with a neutral density filter to less han 1 $\mu$W before coupling it into the ONF. At the output of the ONF, we recombined the probe with a 5 mW LO at a beam splitter, giving a central heterodyne beat frequency of 1 MHz. To guarantee mode overlap between the LO and probe, we aligned them by maximally coupling both beams into a single-mode fiber and matched their polarizations before performing the measurement in free space.

We gain polarization sensitivity by splitting the recombined beam into two arms of equal power and orthogonal polarization at a Wollaston prism. We detected the polarized beams using a balanced detector (Thorlabs PDB450A) with 5 MHz bandwidth, gain of $10^5$, and an output voltage noise {$< 80$ $\mu$V$_{RMS}$}. The signal is digitized with a oscilloscope (Tektronix DPO7054). With this setup, the polarization rotations induced by the torsional modes in the ONF appear as sidebands to the 1 MHz heterodyne beat frequency.
Measurements with ONF 2 have sufficient signal-to-noise ratio that the direct output of the polarimeter with no beating shows the oscillations.

We drive the modes by square-wave-modulating the amplitude of a circularly-polarized 750 nm wavelength laser beam (Coherent 899 titanium-sapphire ring) for ONF 1 or 980 nm beam (distributed feedback laser (DFB)) for ONF-2 using an AOM (``Excitation laser'' in Fig.\ref{fig:ExpSetup}(b)).  Unpolarized continuous wave light can change the resonant frequency of the torsional modes through a thermo-optical effect. To test the effect of a heating beam on the torsional modes, we use a 1064 nm laser (Lumentum NPRO 126-1064). We couple the beam into the ONF and wait several seconds for it to thermalize before taking data. This effect is reversible, affects all the resonant frequencies equally, and does not affect the quality factor of the torsional modes. A variation up to 10 mW of optical power allows tunability of the torsional modes up to 2 kHz with typical thermalization times of a few hundred milliseconds.

\section{Measurements}

We observe the torsional modes by sending a weak (less than 1 $\mu$W) probe through the ONF, and then analyzing its polarization at the output.  The probe is nominally linearly polarized through the length of the waist, owing to the small intrinsic birefringence of the ONF, as verified by Rayleigh scattering diagnostics~\cite{Hoffman2015}.  The thermally-driven modes at room temperature are strong enough to produce a significant modulation of the polarization with good signal-to-noise ratio. This modulation has been identified as a worrisome problem causing parametric heating of trapped atoms \cite{Vetsch2010,Goban2012,Solano2017c,Sayrin2015a,Ostfeldt17}.  

\begin{figure}[h]
\centering
\includegraphics[width=0.7\linewidth]{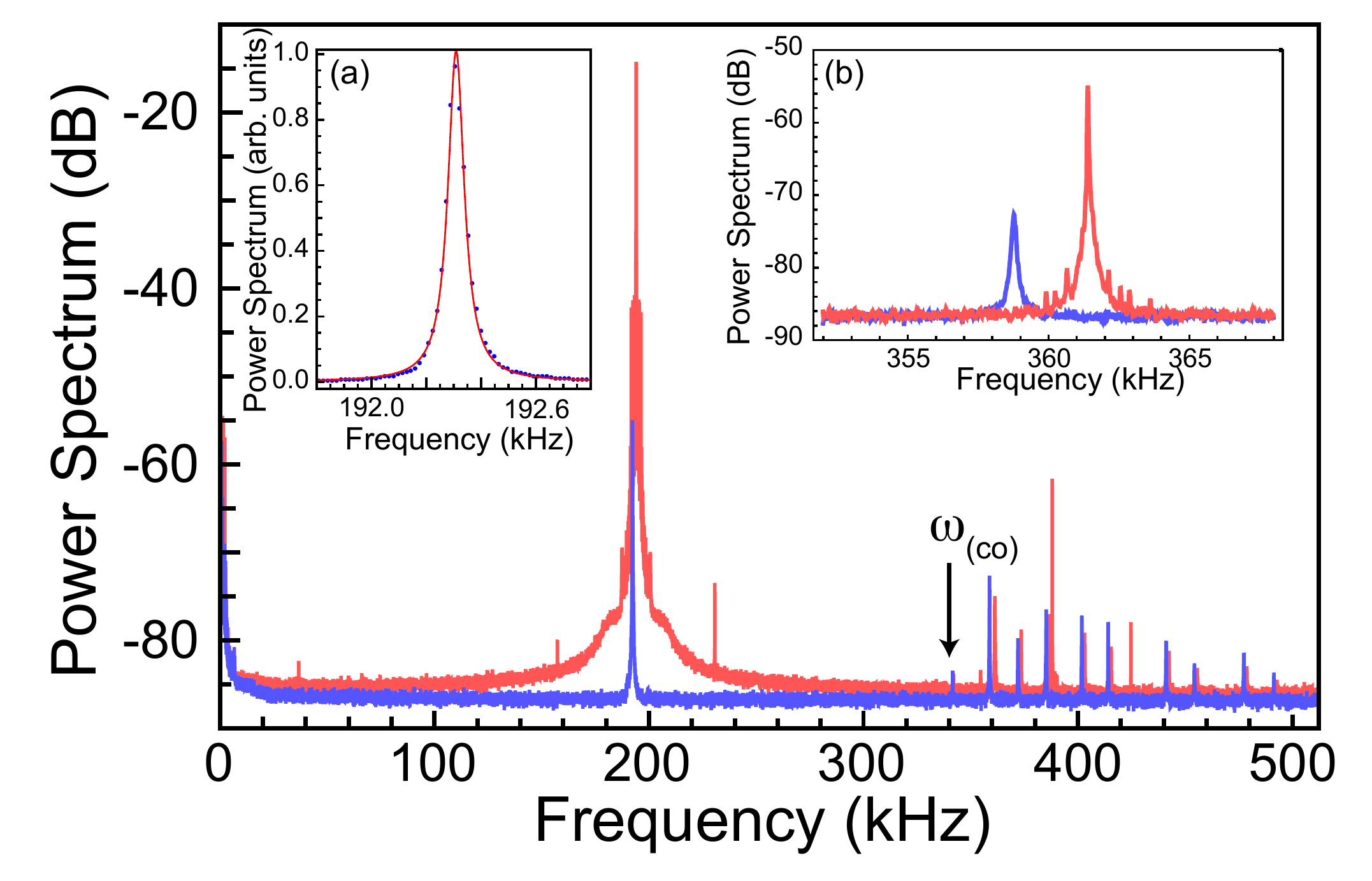}
\caption{FFT of the signal from the polarimeter on ONF-1. (blue) thermally-driven trace. (red) optically-driven at the frequency of the first torsional mode near 200 kHz.  
The arrow indicates the calculated $\omega_{co}$ = 340 kHz using \cite{Wuttke2014}. The vertical scale is logarithmic. Inset (a): Fit (red) of the resonance of the first mode (blue points) to a Lorenztian in linear scale. Inset (b): Change of the frequency of a higher mode from thermal (blue) and driven (red). The vertical scale is logarithmic.}
\label{fig:spectrum}
\end{figure}

The symmetric first torsional mode of an ideal, symmetric ONF should not cause any net polarization rotation at the nanofiber output. 
However, our simulations based on the work of Ref.~\cite{Wuttke2014} have shown that the mechanical evanescent mode that propagates into the taper of the fiber and the asymmetries in the taper are sufficient to create a signal. 
Figure \ref{fig:spectrum} shows the power spectra calculated from the FFT of the polarimeter output for ONF 1 for the thermally driven (blue) and optically driven (red) cases. Below the cutoff frequency $\omega_{co}$ = 340 kHz for ONF 1, the torsional modes are damped by the exponential region of the fiber, and therefore confined to the waist. Above $\omega_{co}$, nodes appear in the exponential section. The modes are no longer confined to the waist and propagate along the fiber~\cite{Wuttke2014}.
Compared to theoretical simulations~\cite{Wuttke2013,Wuttke2014} of the fiber, the measured first torsional frequency and $\omega_{co}$ are consistent with a fiber whose waist is 0.4 mm (6 \%)longer than the expected value for our fiber.

If we drive the first torsional mode of the fiber on resonance the amplitude increases by almost 40 dB (red trace). The inset (a) shows a Lorentzian fit to the first torsional mode with a resulting full-width at half-maximum (FWHM) of 72 Hz at 192.3 kHz producing a Q of 2,671. These results are also consistent with ringdown measurements. We measured similar Q factors for higher-order torsional modes. 
Inset (b) shows the change in resonant frequency for a higher-order torsional mode after resonant excitation, caused by the heating of the excitation laser.  After this amplification, the vibrational (violin) modes become visible.  These violin modes can come not only from the ONF, but from any other mechanical oscillations of mounts, etc, along the fiber. 

As we increase the optical power through the fiber without AM modulation, the frequency of all modes present increases. Though we are not fully sure what causes this effect, we believe it is a geometric effect from thermal expansion of the fiber and thermooptic shifts. We have measured in the first mode of ONF 2 (135.8 KHz)  the linear thermal shift of the resonance to be 0.68 Hz/$\mu$W, for powers up to 800 $\mu$W. 

Light can apply torque to the fiber in two ways.  In a waveguide with strong intrinsic birefringence, as in Ref.~\cite{He2016}, spin angular momentum transfer can occur through the net change in transmitted polarization, even when there is no absorption.  Without birefringence, spin angular momentum can be transferred by being absorbed by or scattered from the medium.  In our ONF, both the intrinsic birefringence and the absorption, and hence torque, are low, but finite.  As in experiments with free microparticles~\cite{He1995,Friese1996}, the direction of transferred spin depends on the polarization of the incident light.  Because the ONF is clamped at its ends, such free rotation is not observable, but by resonantly driving the mechanical system with amplitude-modulated light, the effect of the strain can be become pronounced as Fig.~\ref{fig:spectrum} shows.

\begin{figure}[htb]
\centering
\includegraphics[width=0.70\linewidth]{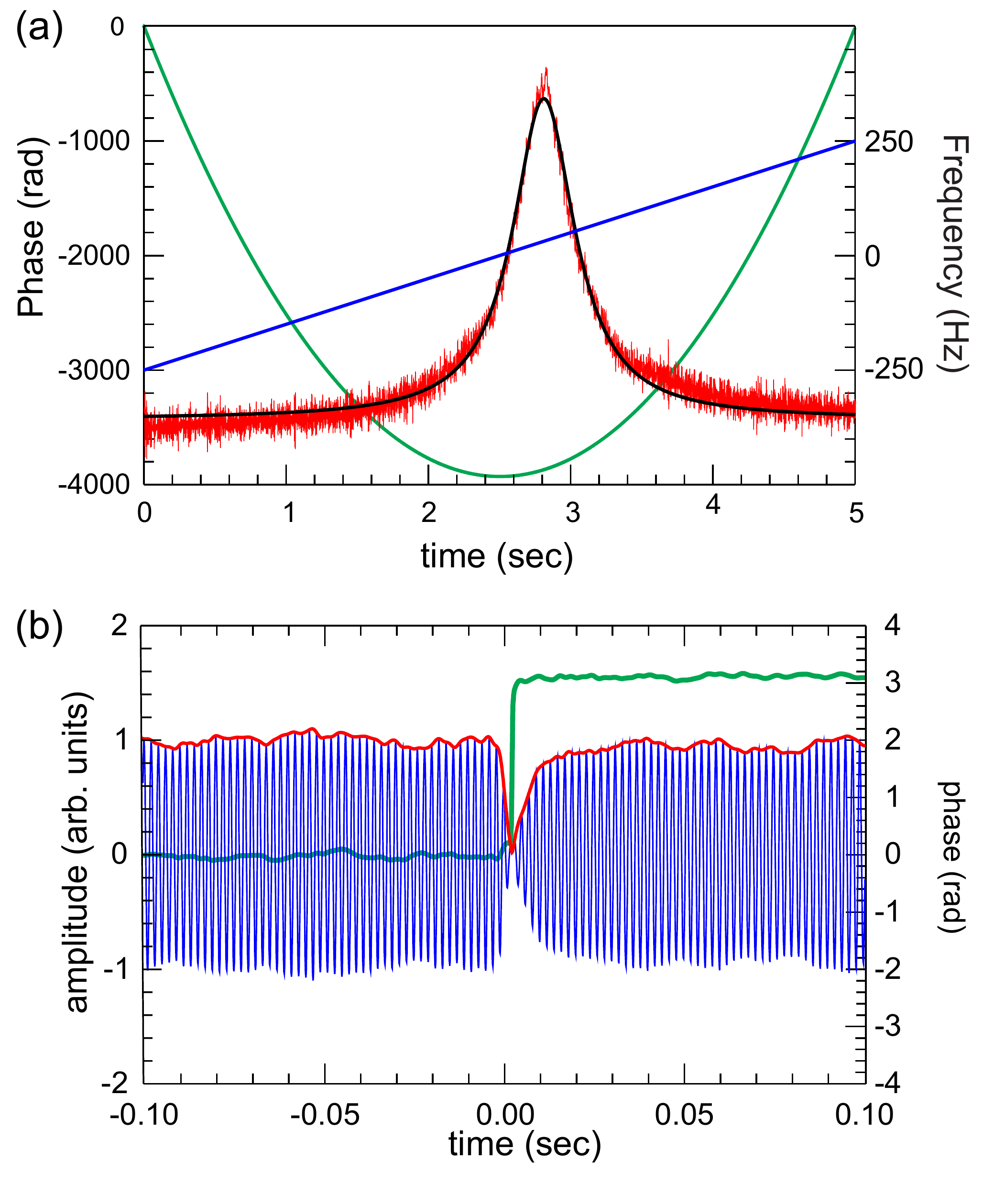}
\caption{Control of the oscillations in ONF 2: a) Response of ONF 2 to 80 $\mu$W 980 nm light across the first 136 kHz torsional mode (FWHM 53 Hz). Demodulated amplitude (red) and phase (green) of the polarimeter signal as the frequency shifts 500 Hz across the first torsional mode with the fit to a Lorenzian peak.  Blue trace: frequency shift as calculated by differentiating the phase with respect to time.  Left axis for the phase, right axis for the frequency ramp. b) Response of the system oscillations on the fundamental mode of ONF 2 (blue) to a $\pi$ phase shift green, showing the rapid change. The red line is the amplitude of the oscillations.}
\label{fig:Fig4}
\end{figure}

While we cannot fully separate the effects due to birefringence and absorption, we believe that the transfer is due primarily to absorption and scattering in the ONF.  To test this, we varied the degree of longitudinal spin in the ONF waist by using a quarter-wave-plate at the fiber input, as follows.  Using Rayleigh scattering~\cite{Hoffman2015}, we could determine input polarization settings that create quasilinear polarization on the ONF waist by minimizing Rayleigh scattered light along a particular viewing direction (We note that in nanophotonic systems, there is always some degree of elliptical polarization that produces transverse spin, but this component does not contribute to the torsional modes~\cite{Lodahl2016}).  In this case, the longitudinal spin is minimized.  Adjusting the input polarization with the quarter-wave-plate increases the longitudinal spin on the waist to its maximum value.  This adjustment increases the amplitude of the resonance at 193 kHz in Fig.~\ref{fig:spectrum} from -55 dB to -15 dB.  For the case of minimized longitudinal spin (quasilinear polarization), the amplitude increased by only 7 dB (not shown).  This small increase for quasilinear polarization could either be due to the slight birefringence in the ONF, with length scale of $>$1 cm~\cite{Fatemi2017}, or to imperfect cancellation of the longitudinal spin.

Since the linear and circular beams are at equal powers, the frequency up-shift induced by heating (2 kHz) is the same for both, corroborating the heat explanation above. The violin modes appear as side-bands to the main mode and do not shift with heat brought by the unmodulated excitation beam.

More control of the oscillations comes from the direct drive of the modes. This allows us to study in detail the Lorentzian response of the resonance. 
Fig.~\ref{fig:Fig4}(a) shows the response of ONF 2 to a chirp (500 Hz) on the modulating frequency around the torsional mode resonance. 
The red trace shows the normalized amplitude of the demodulated output of the polarimeter signal when sweeping the frequency linearly from -250 Hz to +250 Hz relative to the 136390 Hz resonance frequency. The green trace shows the phase of the demodulated polarimeter output. This curve, showing a quadratic dependence of ~4000 rad, can be differentiated to produce the blue trace showing the frequency sweep as a function of time. That the overall system phase matches that of the known driving field means that we have control over the phase of the torsional modes, as expected in a resonantly driven system.  The Lorentzian fit gives a FWHM = 53 Hz.

Fig.~\ref{fig:Fig4}(b) shows the direct polarimeter output in blue for the case when the driving field is suddenly switched by $\pi$.  In a classical driven system, driving the system out of phase will reduce the amplitude as the system responds.  This is clearly shown in the figure.  The red curve is the demodulated amplitude of the polarimeter signal.  The green curve is the demodulated phase, also showing with $\pi$ phase jump.  The system takes a few milliseconds to fully respond, agreeing with our measured Q.

\section{Discussion}
In addition to the fundamental interest in optomechanics based on angular momentum transfer~\cite{Shi2016}, the fact that we can change optical to rotational energy by exciting an ONF with circularly polarized light could make this a very sensitive detector for circular polarization. However, we also have in this system the more exciting realization of polarization modulation transfer that can become a very interesting spectroscopic tool when one combines it with emission from chiral objects \cite{Lodahl2016} around the ONF but coupled through the evanescent field to the mode. 

The phase control of the oscillation is an important feature for driven systems and opens the possibility to many interesting applications in feedback~\cite{Bhattacharya2016}.  This may be a way to decrease the temperature of the mode, but reaching the quantum that would require increasing the $Q$ by a few orders of magnitude.

\section*{Funding Information}
This work has been supported by National Science Foundation of the United States (NSF) (PHY-1307416); NSF Physics Frontier Center at the Joint quantum Institute (PHY-1430094); Army Research Laboratory; and by the Office of the Secretary of Defense through the Quantum Science and Engineering Program.

\section*{Acknowledgments}
We would like to acknowledge the help of S. L. Rolston at the initiation of this research project.


\end{document}